\documentclass[sigconf,screen]{acmart}

\newif\ifacm \acmfalse

\ifacm
\else
\renewcommand\footnotetextcopyrightpermission[1]{} 
\fi

\RequirePackage{url}
\RequirePackage{colortbl}
\RequirePackage{amsmath,amsfonts,amssymb,latexsym}
\usepackage{multirow, makecell}
\RequirePackage{enumitem} \setlist{nolistsep}
\RequirePackage{tabularx,colortbl}
\RequirePackage[skip=\parskip]{caption}
\RequirePackage{voila}
\RequirePackage[textsize=footnotesize]{todonotes}
\DeclareMathOperator{\clo}{clo}

\makeatletter
\g@addto@macro\normalsize{%
  \setlength\abovedisplayskip{0pt}
  \setlength\belowdisplayskip{0pt}
  \setlength\abovedisplayshortskip{0pt}
  \setlength\belowdisplayshortskip{0pt}
  \parskip 1pt plus 1pt
}
\makeatother

\def\SOE{\ensuremath{\sigma\text{OE}}}

\ifacm
\copyrightyear{2018} 
\acmYear{2018} 
\setcopyright{acmlicensed}
\acmConference[SIGIR '18]{The 41st International ACM SIGIR Conference on Research \& Development in Information Retrieval}{July 8--12, 2018}{Ann Arbor, MI, USA}
\acmBooktitle{SIGIR '18: The 41st International ACM SIGIR Conference on Research \& Development in Information Retrieval, July 8--12, 2018, Ann Arbor, MI, USA}
\acmPrice{15.00}
\acmDOI{10.1145/3209978.3210150}
\acmISBN{978-1-4503-5657-2/18/07}
\fi

\begin{document}


\title{New Embedded Representations and Evaluation Protocols \\
for Inferring Transitive Relations}
\author{Sandeep Subramanian}
\affiliation{IIT Bombay}
\author{Soumen Chakrabarti}
\affiliation{IIT Bombay}
\date{}

\begin{abstract}
Beyond word embeddings, continuous representations of knowledge graph (KG) components, such as entities, types and relations, are widely used for entity mention disambiguation, relation inference and deep question answering.  Great strides have been made in modeling general, asymmetric or antisymmetric KG relations using Gaussian, holographic, and complex embeddings.  None of these directly enforce transitivity inherent in the is-instance-of and is-subtype-of relations.  A recent proposal, called order embedding (OE), demands that the vector representing a subtype elementwise dominates the vector representing a supertype.  However, the manner in which such constraints are asserted and evaluated have some limitations.  In this short research note, we make three contributions specific to representing and inferring transitive relations.  First, we propose and justify a significant improvement to the OE loss objective.  Second, we propose a new representation of types as hyper-rectangular regions, that generalize and improve on OE.  Third, we show that some current protocols to evaluate transitive relation inference can be misleading, and offer a sound alternative.  Rather than use black-box deep learning modules off-the-shelf, we develop our training networks using elementary geometric considerations.
\end{abstract}

\maketitle

\section{Introduction}
\label{sec:Intro}

Contemporary information extraction from text, relation inference in knowledge graphs (KGs), and question answering (QA) are informed by continuous representations of words, entities, types and relations.  Faced with the query ``Name \underline{scientists} who played the violin,'' and having collected candidate response entities, a QA system will generally want to verify if a candidate is a scientist.  Testing if $e \in t$ or $t_1 \subseteq t_2$, where $e$ is an entity and $t, t_1, t_2$ are types, is therefore a critical requirement.  Unlike Albert Einstein, lesser-known candidates may not be registered in knowledge graphs, and we may need to assign a confidence score of belongingness to a target type.

A common recipe for inferring general relations between entities is to fit suitable vectors to each of them, and to train a network to input query vectors and predict presence or absence of the probed relationship.  A key question has been whether types merit a special representation, different from the generic devices that represent KG relations, because of their special properties.  Two types may be disjoint, overlapping, or one may contain the other.  Containment is transitive.

Compared to the vast array of entity-relation representations available \citep{BordesUGWY2013TransE, Nickel2013Tensor, NickelRP+2016Holographic,TrouillonWRGB2016Complex, XieMDH2017ITransF}, few proposals exist \citep{VilnisM2014Gauss, VendrovKFU2015OrderEmbeddings,JameelS2016SubspaceEmbedding} for representing types to satisfy their specific requirements.  Of these, only order embedding (OE) by \citet{VendrovKFU2015OrderEmbeddings} directly enforces transitivity by modeling it as elementwise vector dominance.

We make three contributions.  First, we present a significant improvement to the OE loss objective.  Second, we generalize OE to \emph{rectangle} embeddings for types: types and entities are represented by (hyper-)rectangles and points respectively.  Ideally, type rectangles contain subtype rectangles and entity instance points.  Rather than invoke established neural gadgets as black boxes, we introduce constraints and loss functions in a transparent manner, suited to the geometric constraints induced by the task at hand.  Third, we remove a limitation in the training and evaluation protocol of \citet{VendrovKFU2015OrderEmbeddings}, and propose a sound alternative.  Experiments using synsets from the WordNet noun hierarchy (same as \citet{VendrovKFU2015OrderEmbeddings}) show the benefits of our new formulations.  Our code will be available\footnote{\protect\url{https://gitlab.com/soumen.chakrabarti/rectangle}}.

\section{Related work}
\label{sec:Related}

Words and entities\footnote{Also see wiki2vec \protect\url{https://github.com/idio/wiki2vec}} are usually embedded as points or rays from the origin \cite{MikolovSCCD2013word2vec,PenningtonSM2014GloVe,YamadaSTT2017NTEE}.  It is well appreciated that relations need more sophisticated representation \citep{BordesUGWY2013TransE, Nickel2013Tensor, NickelRP+2016Holographic,TrouillonWRGB2016Complex, XieMDH2017ITransF}, but types seem to have fallen by the wayside, in relative terms.  \citet{VilnisM2014Gauss} pioneered a Gaussian density representation for words, to model hypernymy via the asymmetric KL divergence as an inference gadget.  Items $x, y$ are represented by Gaussian densities $g_x, g_y$ (with suitable mean and covariance parameters).  If $x\prec y$ we want low $\KL(g_x\|g_y)$.  Normalized densities with unit mass seem inappropriate for types with diverse population sizes.   \citet{AthiwaratkunW2018ProbOE} have used a thresholded divergence $d_\gamma(x,y) = \max\{0, \KL(g_x\|g_y) - \gamma\}$.
However, modeling asymmetry does not, in itself, enforce transitivity.  Neither is anti-symmetry modeled.  \citet{JameelS2016SubspaceEmbedding} proposed using \emph{subspaces} to represent types.  They do not address type hierarchies or transitive containment.  Recently, \citet{NickelK2017Poincare} introduced an elegant hyperbolic geometry to represent types, but moving away from Euclidean space can complicate the use of such embeddings in downstream applications, in conjunction with conventional word embeddings.  \citet{VendrovKFU2015OrderEmbeddings} proposed a simpler mechanism: embed each type $t$ to vector $\pmb{u}_t \in \R^D$, and, if $t_1 \subseteq t_2$, then require $\pmb{u}_{t_1} \ge \pmb{u}_{t_2}$, where $\ge$ is elementwise.  I.e., $\pmb{u}_{t_1}$ must \emph{dominate} $\pmb{u}_{t_2}$.  OE was found better at modeling hypernymy than Gaussian embeddings.  In OE, types are open cones with infinite volume, which complicates representing various intersections.

\section{\SOE: OE with improved loss objective}

In what follows, we use the partial order $x \prec y$ to unify $e \in t$ and $t_1 \subseteq t_2$ for notational simplicity.  If $x \prec y$, OE required $\pmb{u}_x \ge \pmb{u}_y$.  OE defines $\ell(x,y) = \|\max\{\pmb{0}, \pmb{u}_y-\pmb{u}_x\}\|^2_2$, which is 0 iff $\pmb{u}_y \le \pmb{u}_x$.  Given labeled positive instances $x \prec y$ and negative instances $x \not\prec y$, the overall loss is the sum of two parts:
\begin{align}
\mathcal{L}_+ &= \sum_{x \prec y} \ell_+(x,y) = \sum_{x \prec y} \ell(x,y) \\
\mathcal{L}_- &= \sum_{x \not\prec y} \ell_-(x,y) =
\sum_{x \not\prec y} \max\{0, \alpha - \ell(x,y)\},
\end{align}
where $\alpha$ is a tuned additive margin.  The intuition is that when $x \not\prec y$, we want $\ell(x,y) \ge \alpha$.  There are two limitations to the above loss definitions.  First, $\|\cdots\|_2^2$ is too sensitive to outliers.  This is readily remedied by redefining $\ell(x,y)$ using L1 norm, as
\begin{align}
\|\max\{\pmb{0},\pmb{u}_y-\pmb{u}_x\}\|_1 = \sum_{d=1}^D[u_{y,d} - u_{x,d}]_+,
\end{align}
where $[\bullet]_+ = \max\{0,\bullet\}$ is the hinge/ReLU operator.  But the semantics of $\ell_-$ are wrong: we are needlessly encouraging \emph{all} dimensions to violate dominance, whereas violation in just \emph{one} dimension would have been enough.

Specifically, for $x \not\prec y$, loss should be zero if $u_{x,d} < u_{y,d}$ for any $d\in[1,D]$.  Accordingly, we redefine
\begin{align}
\ell_-(x,y) &= \min_{d\in[1,D]} [u_{x,d} - u_{y,d}]_+,
\end{align}
so that the loss is zero if dominance fails in at least one dimension.  To balance this $L_\infty$ form in case of positive instances, we redefine
\begin{align}
\ell_+(x,y) &= \max_{d\in[1,D]} [u_{y,d} - u_{x,d}]_+,
\end{align}
so that the loss is zero only if dominance holds in all dimensions.

The unbounded hinge losses above mean a few outliers can hijack the aggregate losses $\mathcal{L}_+$ and $\mathcal{L}_-$.  Moreover, the absence of a SVM-like geometric margin (as distinct from the \emph{loss} margin $\alpha$ above) also complicates separating $\prec$ and $\not\prec$ cases confidently.  Our final design introduces a nonlinearity (sigmoid function) to normalize per-instance losses, additive margin $\Delta$ and a standard stiffness hyperparameter $\psi$.
\begin{align}
\ell_+(x,y) &= \sigma\left(\psi \max_{d\in[1,D]} [u_{y,d} + \Delta - u_{x,d}]_+
\right) - 1/2 \\
\ell_-(x,y) &= \sigma\left(\psi \min_{d\in[1,D]} [u_{x,d} + \Delta - u_{y,d}]_+
\right) - 1/2.
\end{align}
(Obviously the `$-1/2$' terms are immaterial for optimization, but bring the loss expression to zero when there are no constraint violations.)

\section{Rectangle embeddings}
\label{sec:Rectangle}

Despite its novelty and elegance, OE has some conceptual limitations.  A type $t$ with embedding $\pmb{u}_t$ is the infinite axis-aligned open convex cone $\{\pmb{p}: \pmb{p} \ge \pmb{u}_t\}$ with its apex at $\pmb{u}_t$.  Thus, types cannot ``turn off'' dimensions,  all pairs of types intersect (although the intersection may be unpopulated), and all types have the same infinite measure, irrespective of their training population sizes.

We propose to represent each type by a \emph{hyper-rectangle} (hereafter, just `rectangle'), a natural generalization of OE cones.  A rectangle is convex, bounded and can have collapsed dimensions (i.e., with zero width).  Obviously, rectangles can be positioned to be disjoint, and their sizes can give some indication of the number of known instances of corresponding types.  Containment of one rectangle in another is transitive by construction, just like OE.  Entities remain represented as points (or infinitesimal rectangles for uniform notation).

Each type or entity $x$ is represented by a base vector $\pmb{u}_x$, as well as a nonnegative \emph{width} vector $\pmb{v}_x \in \R^D_+$, so that in dimension $d$, the rectangle has extent $[u_{x,d}, u_{x,d} + v_{x,d}]$.  Informally, the rectangle representing $x$ is bounded by ``lower left corner'' $\pmb{u}_x$ and ``upper right corner'' $\pmb{u}_x + \pmb{v}_x$.  For entities, $\pmb{v}_x \equiv \pmb{0}$.  For types, $\pmb{v}_x$ are regularized with a L2 penalty.  The rectangles are allowed to float around freely, so $\pmb{u}_x$ are not regularized.

If $x \in y$ or $x \subseteq y$, the rectangle representing $x$ must be contained in the rectangle representing~$y$.  Let the violation in the $d$th dimension be
\begin{align}
\delta_+(x,y;d)= \max\{& [u_{y,d} - u_{x,d}]_+, \\
& [u_{x,d} + v_{x,d} - u_{y,d} - v_{y,d}]_+ \} \notag 
\end{align}
Then the loss expression for positive instances is
\begin{align}
\mathcal{L}_+ &= \sum_{x \prec y} \max_{d\in[1,D]} [ \delta_+(x,y;d) ]_+.
\end{align}
This ensures that the loss is proportional to the largest violating margin and that the loss is zero if the rectangle of $x$ is contained in the rectangle of~$y$. Analogously, we define 
\begin{align}
\delta_-(x,y;d) = \min\{ & [u_{x,d} - u_{y,d}]_+, \\
& [u_{y,d} + v_{y,d} - u_{x,d} - v_{x,d}]_+ \}  \notag
\end{align}
\begin{align}
\text{and} \; \mathcal{L}_-
&= \sum_{x \not\prec y} \min_{d\in[1,D]} [\delta_-(x,y;d)]_+.
\end{align}
As in \SOE, we can add margin, stiffness, and nonlinearity to rectangles, 
and get
\begin{align}
\delta_+(x,y;d)= \max\{ & [u_{y,d} + \Delta - u_{x,d}]_+, \\
& [u_{x,d} + v_{x,d} + \Delta - u_{y,d} - v_{y,d}]_+ \} \notag \\
\delta_-(x,y;d) = \min\{ & [u_{x,d} + \Delta - u_{y,d}]_+, \\
& [u_{y,d} + v_{y,d} + \Delta - u_{x,d} - v_{x,d}]_+ \}  \notag
\end{align}
\begin{align}
\mathcal{L}_+ = \sum_{x\prec y} \ell_+(x,y)
&= \sum_{x\prec y}
\sigma\left( \psi \max_{d\in[1,D]} \delta_+(x,y;d) \right) \\
\mathcal{L}_- = \sum_{x\not\prec y} \ell_-(x,y)
&= \sum_{x\not\prec y}
\sigma\left( \psi \min_{d\in[1,D]} \delta_-(x,y;d) \right).
\end{align}

\section{Training and evaluation protocols}
\label{sec:Eval}

Because the training and evaluation instances are tuple samples from a single (partially observed) partial order, great care is needed in designing the training, development and testing folds.  To use unambiguous short subscripts, we call them learn, dev and eval folds, each with positive and negative instances $L_+, L_-, D_+, D_-, E_+, E_-$.  Let $T$ be the raw set of tuples ($x\in t$ or $t_1 \subseteq t_2$).  The transitive closure (TC) of $T$, denoted $\clo(T)$, includes all tuples implied by $T$ via transitivity.

\subsection{OE protocol}

\citet{VendrovKFU2015OrderEmbeddings} followed this protocol:
\begin{enumerate}
\item Compute $\clo(T)$.
\item Sample positive eval fold $E_+ \subset \clo(T)$.
\item Sample positive learn fold $L_+ \subset \clo(T) \!\setminus\! E_+$.
\item Sample positive dev fold $D_+ \subset \clo(T) \setminus (E_+ \cup L_+)$.
\item Generate negative eval, learning and dev folds, $E_-, L_-$ \& $D_-$ (see below).
\item Return $L_+, L_-, D_+, D_-, E_+, E_-$.
\end{enumerate}
A negative tuple is generated by taking a positive tuple $(x,y)$ and perturbing either of them randomly to $(x,y')$ or $(x',y)$, where $x', y'$ are sampled uniformly at random.  In OE negative folds were the same size as positive folds.

The WordNet \cite{MillerBFGMT1993wordnet5} hypernymy data set used by \citet{VendrovKFU2015OrderEmbeddings} has $|T|=82115$ and $|\clo(T)|=838073$.  $E_+$ and $D_+$, sampled from $\clo(T)$, had only 4000 tuples each.  \emph{All remaining tuples} were in the learn fold.  \citet{VendrovKFU2015OrderEmbeddings} freely admit that ``the majority of test set edges can be inferred simply by applying transitivity, giving [them] a strong baseline.''  They reported that the TC baseline gave a 0/1 accuracy of 88.2\%, Gaussian embeddings \citep{VilnisM2014Gauss} was at 86.6\%, and OE at 90.6\%.

\begin{figure}[ht]
\centering\includegraphics{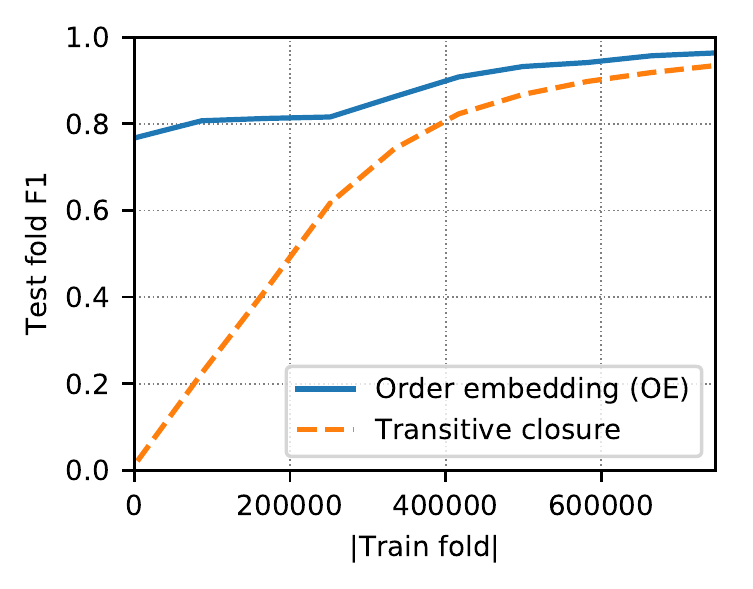}
\caption{A large fraction of test instances can be inferred by simply computing the transitive closure of the training fold in the OE protocol.}
\label{fig:TcGoodEnough}
\end{figure}

Instead of 0/1 accuracy, Figure~\ref{fig:TcGoodEnough} shows the more robust F1 score on test instances achieved by transitive closure and OE, as the size of training data is varied.  \citet{VendrovKFU2015OrderEmbeddings} reported accuracy near the right  end of the scale, where OE has little to offer beyond TC.  In fact, OE does show significant lift beyond TC when training data is scarce.  As we shall see, even with ample training data, \SOE\ and rectangle embeddings improve on OE.

\subsection{Sanitized OE protocol}

Clearly, evaluation results must be reported separately for instances that cannot be trivially inferred via TC, where the algorithm needs discover a suitable geometry from the combinatorial structure of $\clo(T)$ beyond mere reachability.  To this end, we propose the following sanitized protocol.
\begin{enumerate}
\item Sample positive learn fold $L_+ \subset \clo(T)$.
\item Negative learn fold $L_-$ of size $|L_+|$ is generated by repeating as needed:
	\begin{enumerate}
    \item Sample $(u, v) \in L_+$ uniformly.
    \item Perturb one of $u$ or $v$ to get $(u', v')$.
    \item If $(u', v') \in \clo(T)$, discard.
    \end{enumerate}
\item Sample positive dev fold $D_+ \subset (\clo(T)\setminus\clo(L_+))$.
\item Discard $(u, v)$ from $D_+$ if $u$ or $v$ not found in $L_+ \cup L_-$ (explained below).
\item Sample positive eval fold $E_+ \subset (\clo(T)\setminus(\clo(L_+) \cup \clo(D_+)))$.
\item Discard elements from $E_+$ using the same protocol used to discard elements from~$D_+$.
\item Generate negative dev and eval folds, $D_-$ and $E_-$, using the same protocol used to generate $L_-$ from~$L_+$.
\end{enumerate}
An entity or type never encountered in the learn fold cannot be embedded meaningfully (unless corpus information is harnessed, see Section~\ref{sec:Future}), so it is pointless to include in dev or eval folds instances that contain such entities or types.  Such sampled instances are discarded.  To fill folds up to desired sizes, we repeatedly sample pairs until we can retain enough instances.

\section{Experiments}
\label{sec:Expt}

\paragraph{Data set:}
We prepare our data set similar to \citet{VendrovKFU2015OrderEmbeddings}.
WordNet \citep{MillerBFGMT1993wordnet5} gives 82115 (hypernym, hyponym) pairs which we use as directed edges to construct our KG.  The WordNet noun hierarchy is prepared by experts, and is also at the heart of other type systems \citep{SuchanekKW2007YAGO,MurtyVVM2017TypeNet} used in KG completion and information extraction.  We augment the KG by computing its transitive closure, which increases the edge count to 838073.  Then we use the two protocols in Section~\ref{sec:Eval} to create training, dev and test folds.  The sanitized protocol produces 679241 positive and 679241 negative training instances, 4393 positive and 4393 negative dev instances, and 4316 positive and 4316 negative test instances.  These sizes are close to those of \citet{VendrovKFU2015OrderEmbeddings}.

\paragraph{Code and hyperparameter details:}
OE and our enhancements, \SOE\ and rectangle embeddings, were coded in Tensorflow with Adam's optimizer.  Hyperparameters, such as batch size (500), initial learning rate (0.1), margin $\Delta$ and stiffness $\psi$, were tuned using the dev fold.  Optimization was stopped if the loss on the dev fold did not improve more than 0.1\%  for 20 consecutive epochs.  All types and entities were embedded to $D=50$ dimensions.

\paragraph{Results:}
\citet{VendrovKFU2015OrderEmbeddings} reported only microaveraged 0/1 accuracy (`Acc').  Here we also report average precision (AP), recall (R), precision (P) and F1 score, thus covering both ranking and set-retrieval objectives.  AP and R-P curves are obtained by ordering test instances by the raw score given to them by OE, \SOE, and rectangle embeddings.  Table~\ref{tab:wordnet} compares the three systems after using the two sampling protocols to generate folds.

\begin{table}[ht]
\newcolumntype{Y}{>{\centering\arraybackslash}X}
\def\tcy{\cellcolor{yellow!50}}
\def\tcr{\cellcolor{red!30}}
\def\tcp{\cellcolor{green!15}}
\def\tcg{\cellcolor{green!35}}
\centering
\begin{tabularx}{\columnwidth}{ |c| *{6}{Y|}}
\hline
    & \multicolumn{3}{c|}{OE protocol} & \multicolumn{3}{c|}{Sanitized OE protocol} \\ \hline
    & OE    & \SOE  & Rect    & OE      & \SOE  & Rect      \\ \hline
Acc & \tcy 0.922 & 0.921 & 0.926   & \tcr 0.574 & \tcp 0.742   & \tcg 0.767 \\ \hline
AP  & 1     & 1     & 1       & 0.977    & 0.969 & 0.986     \\ \hline
P   & 0.994 & 0.915 & 0.973   & 0.987    & 0.925 & 0.983     \\ \hline
R   & 0.850 & 0.929 & 0.877   & 0.151    & 0.527 & 0.544     \\ \hline
F1  & \tcy 0.916 & 0.922 & 0.923 & \tcr 0.262 & \tcp 0.671 & \tcg 0.700    \\ \hline
\end{tabularx}
\caption{Performance of OE, \SOE, and rectangle embeddings under the OE protocol and the sanitized protocol, on the WordNet hypernymy relation.}
\label{tab:wordnet}
\end{table}

It is immediately visible that absolute performance numbers are very high under the original OE protocol, for reasons made clear earlier.  As soon as the OE protocol is replaced by the sanitized protocol, no system is given any credit for computing transitive closure.  The 0/1 accuracy of OE drops from 0.922 to 0.574.  F1 score drops even more drastically from 0.916 to 0.262.  In contrast, \SOE\ and rectangle embeddings fare better overall, with rectangle embeddings improving beyond \SOE.

\begin{figure}[ht]
\centering\includegraphics{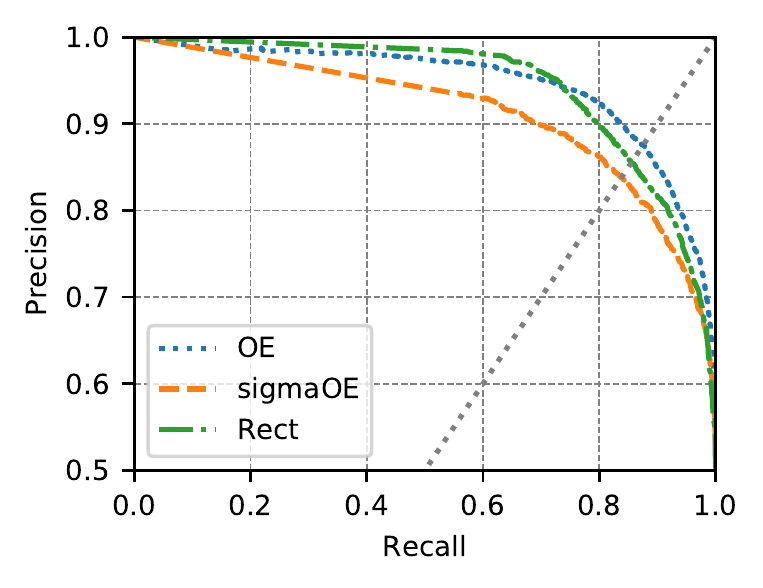}
\caption{Recall-precision profiles on WordNet.}
\label{fig:wordnet}
\end{figure}

Whereas \SOE\ and rectangle embeddings improve on OE at the task of set retrieval, their ranking abilities are slightly different.   Figure~\ref{fig:wordnet} shows that \SOE\ is inferior at ranking to both OE and rectangle embeddings.  Rectangle embeddings have the best precision profile at low recall.  Modifying our code to use ranking-oriented loss functions \citep{CaoQLTL2007ListNet} may address ranking applications better.

\section{Concluding remarks}
\label{sec:Future}

Here we have addressed the problem of completing $e\in t$ and $t_1 \subseteq t_2$ relations starting from an incomplete KG, but without corpus support.  For out-of-vocabulary (not seen during training) entities, mention contexts in a corpus are vital typing clues \citep{LingW2012FineType,YaghoobzadehS2015FineType,ShimaokaSIR2016FineType}.  We plan to integrate context (word) embeddings with order and rectangle embeddings.  It would be of interest to see how our refined loss objectives and testing protocols compare with other corpus-based methods \citep{ChangWVM2017Hypernym,YamaneTYMS2016HypernymGeneration}.

\paragraph*{\bfseries Acknowledgment:}
Thanks to Aditya Kusupati and Anand Dhoot for helpful discussions, and nVidia for a GPU grant.

\bibliographystyle{ACM-Reference-Format}
\bibliography{voila}

\end{document}